\documentclass[11pt,a4paper,fleqn]{article}
\usepackage{amsmath}
\usepackage{t1enc}
\usepackage[latin1]{inputenc}
\usepackage[english]{babel}
\usepackage{graphicx}
\input{psfig.sty}  

\begin{document}
\title{X-matter induced cosmological scenarios in the relativistic theory of gravity}                    

\author{V. L. Kalashnikov}
   
\maketitle
\noindent
\begin{center}
Institut f\"ur Photonik, TU Wien, Gusshausstr. 27/387, A-1040 Vienna, Austria, tel:+43-1-58801-38723, 
fax:+43-1-58801-38799,\\
\noindent
e-mail:vladimir.kalashnikov@tuwien.ac.at
\end{center}
           
\begin{abstract}
It was shown, that the presence of the so-called X-matter with an equation 
of state, which lies between limits of the strong and weak energy conditions, allows 
the variety of the cosmological scenarios in the relativistic theory of 
gravity. In spite of the fixed negative sign of the cosmological term in the 
field equations with massive graviton, it is possible to obtain the solutions with 
accelerated and complicated loitering expansion of the universe. The 
numerical estimation of the universe's age agrees with the modern
observational data if the upper limit of the graviton's mass is $10^{-71}$ \textit{g}
\end{abstract}

\section{Introduction}
The so-called relativistic theory of gravity (RTG), which was inspired by
A.A. Logunov with co-authors \cite{1, 2}, is the alternative to Einstein's
general theory of relativity (GR) and differs from it in the cardinal point:
the gravitation is considered as the tensor field of the Faraday-Maxwell's type
in the flat Minkowski spacetime. The existence of the flat reference
background metric in RTG allows of the unambiguity and clarity of the
obtained solutions, which are similar to ones in GR but to have the property
of the trivial extendibility and topological simplicity. The lasts exclude
the existence of the singularity, black holes, wormholes, topological defects etc.
The structure of the theory is tougher due to existence of the additional field equation
on background metric, which is analog of the harmonic coordinate equation in
GR and is introduced in order to exclude the spins 1 and 0 for gravitons.
The additional advantages of RTG are the existence of the conservation lows
and the possibility of the canonical quantization.

All experimental consequences of RTG are the same as ones in GR for
experiments in solar system. But derived in \cite{3, 4} cosmological scenario
differs essentially: RTG permits only flat global spacetime
universe, critical density of the matter and recollapse of universe in the
case of nonzero mass of graviton. The predictions of the universe's flatness
as well as the absence of the cosmological singularity in the case of
nonzero mass of graviton are the advantages of the theory. But the
decelerated character of the cosmological expansion and critical density of
matter don't agree with the modern observational data.

The modern estimations of the basic cosmological parameters are presented in the
Table 1 ($H_{0}$ is the Hubble constant in \textit{km s}$^{ - 1}$\textit{ Mps}$^{ - 1}$,
\textit{$\Omega $}$_{K, M, X}$ are
the normalized density parameters for the curvature, matter and cosmological
constant, respectively, $\tau_{0}$ is the minimal estimation of the universe's
age corresponding to age of oldest globular clusters in \textit{Gyr}, $w$ is the parameter
of the equation of state for the dominant energy (see below)).

\begin{table}
\caption{Cosmological parameters \cite{5}}
\begin{center}
\begin{tabular}{|c|c|}
\hline
 cosmological parameters & observational data\\
\hline
        $H_{0}$& 68$\pm $6 \\
        $\Omega_{K}$ & 1.1$\pm $0.12 \\
			 $\Omega_{M}$ & 0.35$\pm $0.1 \\
			 $\Omega_{X}$ & 0.65$\pm $0.15 \\
        $\tau_{0}$& 12.7$\pm $3 \\
        $w$ & $\leq -0.6$\\
\hline
\end{tabular}
\end{center}
\end{table}

The domination
of \textit{$\Omega $}$_{X}$ provides the accelerated expansion of the universe because of the
acceleration parameter ${\left. {{\frac{{\ddot {a}}}{{aH^{2}}}}}
\right|}_{0} = \Omega _{X} - {\frac{{\Omega _{M}} }{{2}}} > 0$ ($a$ is the
scaling factor, $H$ is the Hubble function, dote denotes the derivative on the
time, the values are considered at the present moment) \cite{6, 7}.

In this work we demonstrate the possibility of variety of cosmological scenarios in RTG with
the so-called "dark energy" (X-matter) term, that adjusts the RTG predictions with the modern
astronomical data.

\section{Model}
We can not introduce the cosmological constant $\Lambda$ with a positive sign in RTG aimed
to describe the "repulsive" action of the vacuum without modification of the field equations. 
Indeed the field equations containing
$\Lambda$ and terms describing the nonzero mass of graviton $m$ have to produce a zero
curvature of the efficient Riemannian spacetime, when the tensor of the energy-momentum
of matter $T^{m}_{n} = 0$ \cite{2}. That results in $\Lambda = -m^{2}$ and, as consequence,
"attracting" action of the vacuum (or zero cosmological constant for the massless graviton).
As it was shown in \cite{8}, there is the simple possibility to modification of the basic
field equations due to introduction of the massive scalar field that results in the inflation
scenario and accelerated expansion of the universe. Such approach is elegant, but here we will
consider the pure phenomenological alternative taking into consideration "dark energy" (X-matter), which
does not cluster on the small scales, is weakly coupled to ordinary matter and has phenomenological equation of state
$p_{X} = w_{X} \rho_{X}$ ($p$ is the pressure, $\rho$ is the density), where $w_{X}$ lies
between 0 (dust of the "usual" matter) and -1 (true cosmological constant in GR). There exist
the different candidates for such X-matter. The consideration of these models is not matter
of this work (for overview see \cite{9}) and we restrict oneself to the phenomenological
equation of state with constant negative $w_{X}$.

The Logunov's field equations are:

\begin{equation}\label{1}
G_n^m  - \frac{{m^2 }}{2}(\delta _n^m  + g^{mk} \gamma _{kn}  - 
\frac{1}{2}\delta _n^m g^{pk} \gamma _{pk} ) =  - 8\pi T_n^m,
\end{equation}

\begin{equation}\label{2}
D_m \tilde g^{mn}  = 0,
\end{equation}

\noindent
where $G_{n}^{m}$ is the Einstein's tensor defined on the efficient Riemannian spacetime
with metrics $g^{mn}$, $\gamma^{mn}$ is the metrics of the flat background Minkowski spacetime,
$D_{m}$ is the covariant derivative on the background spacetime,
$\tilde g^{mn} = \sqrt { - g} g^{mn}$, $c = G = \hbar = 1$. The choice of signs in Eq. \ref{1}
is defined by the use in the calculations of the computer algebra system Maple 6 \cite{10}.

Under the assumption of homogeneity and isotropy of the efficient Riemannian spacetime
(the background spacetime has a Galilean metric in our case), its interval has the following
form in the spherical coordinates \cite{2, 11}:

\begin{equation}\label{3}
ds^2  = d\tau ^2  - a(\tau )^2 \left[ {dr^2  + r^2 \left( {d\theta ^2  + 
\sin (\theta )^2 d\phi ^2 } \right)} \right],
\end{equation}

\noindent
where $\tau$ is the proper time, $a(\tau)$ is the scaling factor as function of the proper time.
Hence from Eqs. (\ref{1} - \ref{3}) we have the evolutional equations:

\begin{equation}\label{4}
\left( {\frac{{\dot a}}{a}} \right)^2  = \frac{8}{3}\pi \rho (\tau ) - 
\frac{1}{6}\frac{{m^2 \left( {a^2  - 1} \right)^2 \left( {a^2  + \frac{1}{2}} \right)}}{{a^6 }},
\end{equation}

\begin{equation}\label{5}
\frac{{\ddot a}}{a} =  - \frac{4}{3}\pi \left( {3p(\tau ) + \rho (\tau )} \right) - 
\frac{1}{6}m^2 \left( {1 - \frac{1}{{a^6 }}} \right).
\end{equation}

\noindent
Then, from the phenomenological equation of state $p = w \rho$ and covariant conservation low
$\nabla _n T_m^n  = 0$ (here $\nabla_n$ is the covariant derivative on Riemannian spacetime), we can
obtain from Eqs. (\ref{4}, \ref{5}):

\begin{eqnarray}\label{6}
\left( {\frac{{da(t)}}{{dt}}} \right)^2  = \frac{{1 - \Omega _X  - \Omega _R }}{{a(t)}} + 
\frac{{\Omega _X }}{{a(t)^{1 + 3w_X} }} +\\ \nonumber
\frac{{\Omega _R }}{{a(t)^2 }} -
\Omega _G \left( {a(t)^2  + \frac{1}{{2a(t)^4 }} - \frac{3}{{2a(t)^2 }}} \right),
\end{eqnarray}

\begin{eqnarray}\label{7}
\frac{{d^2 a(t)}}{{dt^2 }} =  - \frac{{1 - \Omega _X  - \Omega _R }}{{2a(t)^2 }} - 
\frac{{\left( {1 + 3w_X } \right)}}{{2a(t)^{2 + 3w_X } }} -\\ \nonumber
\frac{{\Omega _R }}{{a(t)^3 }} - \Omega _G \left( {a(t) - \frac{1}{{a(t)^5 }}} \right),
\end{eqnarray}

\noindent
where ${t = H}_{\rm 0} (\tau  - \tau _0 )$, $a(t)$ is the
scaling factor normalized to its present value, $\Omega _R  = \frac{{8\pi \rho _R }}{{3H_0^2 }}$ is
the density parameter for the radiation (including relativistic neutrinos) with
$w_R = 1/3$, $\Omega _X  = \frac{{8\pi \rho _X }}{{3H_0^2 }}$ is the density parameter for
the X-matter with unknown negative $w_X$, $\Omega _G  = \frac{{m^2 }}{{6H_0^2 }}$ is the density
parameter for the gravitons. For the "usual" matter with density parameter
$\Omega _M  = \frac{{8\pi \rho _M }}{{3H_0^2 }}$ we supposed $w_M = 0$. Also, we used the cosmic
sum rule $\Omega _M  + \Omega _R  + \Omega _X  = 1$ and took into consideration that the graviton's
term doesn't contribute to this rule. As the estimation of $w_X$ can be used the extremal values of
the so-called weak energy condition $\rho + p$ $\geq$0 and strong energy condition $\rho + 3 p$ $\geq$0 \cite{12}.

\section{Discussion}

The dynamical properties of the system (\ref{6}, \ref{7}) can be found by the search of the zeros of
the right-hand sides of these equations. Then from the Eq. \ref{6} we have the following polynomial
equation for the minimal $a_{min}$ and maximal $a_{max}$ scaling factors of the universe:

\begin{eqnarray}\label{8}
2(\Omega _X a^{3 - 3w}  - \Omega _G a^6 ) + 2(1 - \Omega _R  - \Omega _X )a^3\\ \nonumber
  + (2\Omega _r + 3\Omega _G )a^2  - \Omega _G  = 0
\end{eqnarray}

The dependences of $a_{min}$ and $a_{max}$ on $\Omega_G$ are shown in Fig. 1 for fixed
$\Omega_R$ $\approx$ $9.1 \cdot 10^{-5}$ \cite{12}. As it is known \cite{2,3,4}, $a_{min} \ne 0$
for nonzero $m$ (curve 1). For $a_{min}< 10^{-4} \approx 1/\Omega_R $ (that is the approximate
value of scaling factor at the end of the radiation epoch) the dependence of the minimal scaling
factor on $\Omega_X$ (and $\Omega_M$) is negligible and we can write for this value:

\begin{equation}\label{9}
a_{\min }  \approx \sqrt {\frac{{\Omega _G }}{{2\Omega _R  + 3\Omega _G }}}.
\end{equation}

 \begin{figure}
   \begin{center}
   \begin{tabular}{c}
   \psfig{figure=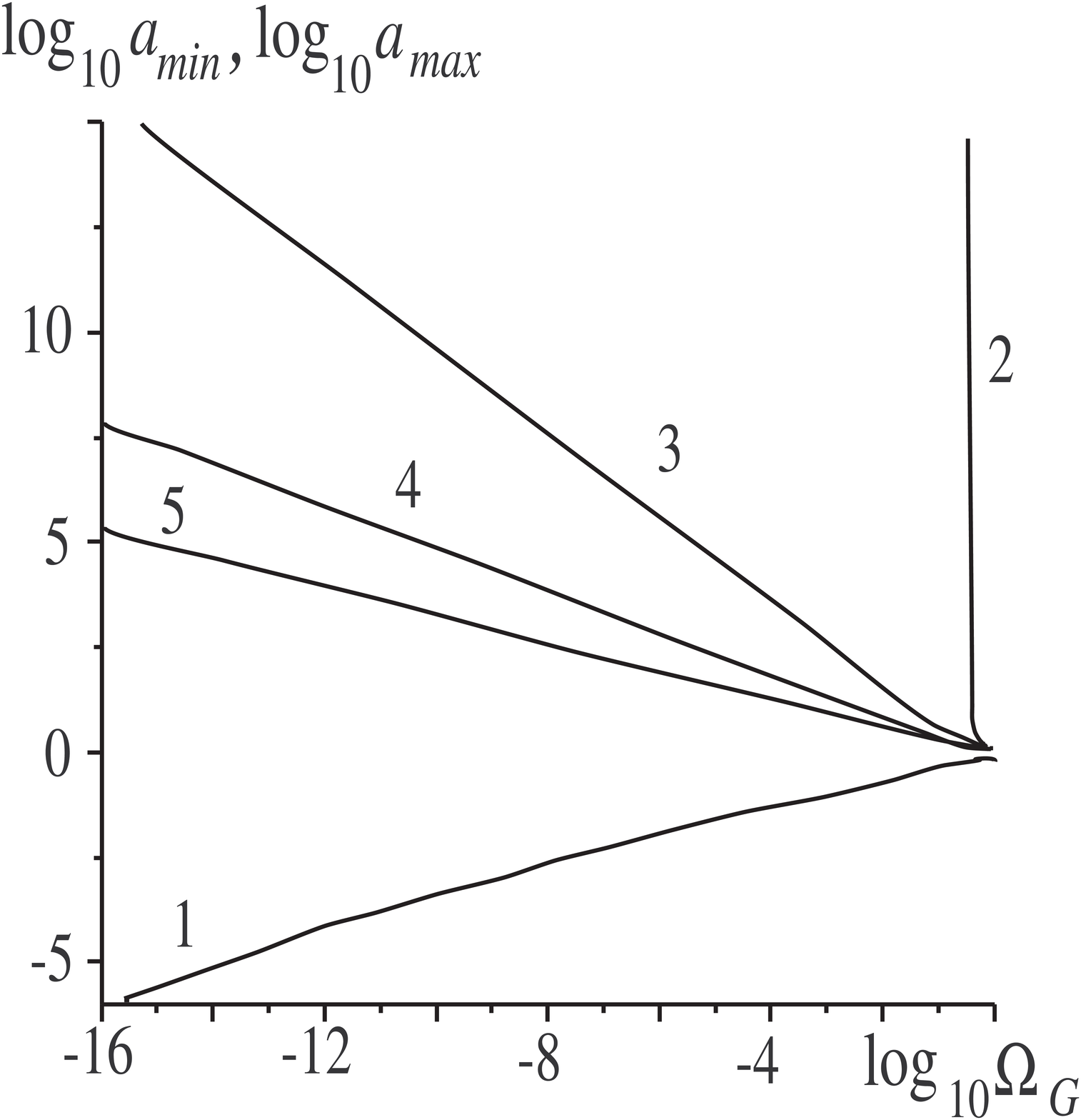,height=7cm} 
   \end{tabular}
   \end{center}
   {Fig. 1. The logarithms of the minimal (curve 1) and maximal (curves 2-5) scaling factors in
        the dependence on the logarithm of the graviton's critical density. $\Omega_X$ = 0.65 (1-4),
        0 (5), $w$ = -1 (2), -2/3 (3), -1/3 (4)} 
   \end{figure} 

\noindent
But the original feature of our model is the strong dependence of $a_{max}$ on the presence and
state of X-matter. If its equation of state corresponds to the pure $\Lambda$-term ($w$ = -1),
there exist the infinitely expanding solutions of Eqs. (\ref{6}, \ref{7}) for $\Omega_G < \Omega_X$
(curve 2, see also the asymptotically growing term in Eq. \ref{8}). The transition to more "moderate"
$w$ (curves 3, 4) results in the recollapsing behavior of the universe with decreasing $a_{max}$ as
result of $\left| w \right|$ decrease. The value of $a_{max}$ in the absence of X-matter is
shown by curve 5. For fixed $w$ the growth of $\Omega_X$ from 0.5 to 0.8 (see Table 1) increases
$a_{max}$ but this dependence is reduced by $\left| w \right|$ decrease.

Accelerated (decelerated) character of the universe's expansion is defined by the sign of the polynomial,
which is the modified right-hand side of Eq. \ref{7}:

\begin{eqnarray}\label{10}
\Omega _X a^{3 - 3w} (1 + 3w) + 2\Omega _G a^6  + (1 - \Omega _X  - \Omega _R )a^3\\ \nonumber
  + 2\Omega _R a^2  - 2\Omega _G
\end{eqnarray}

\noindent
The roots of this polynomial give the values of the critical scaling factor $a_{cr}$ (Fig. 2)
determining the transition from the accelerated (decelerated) to the decelerated (accelerated) expansion.

 \begin{figure}
   \begin{center}
   \begin{tabular}{c}
   \psfig{figure=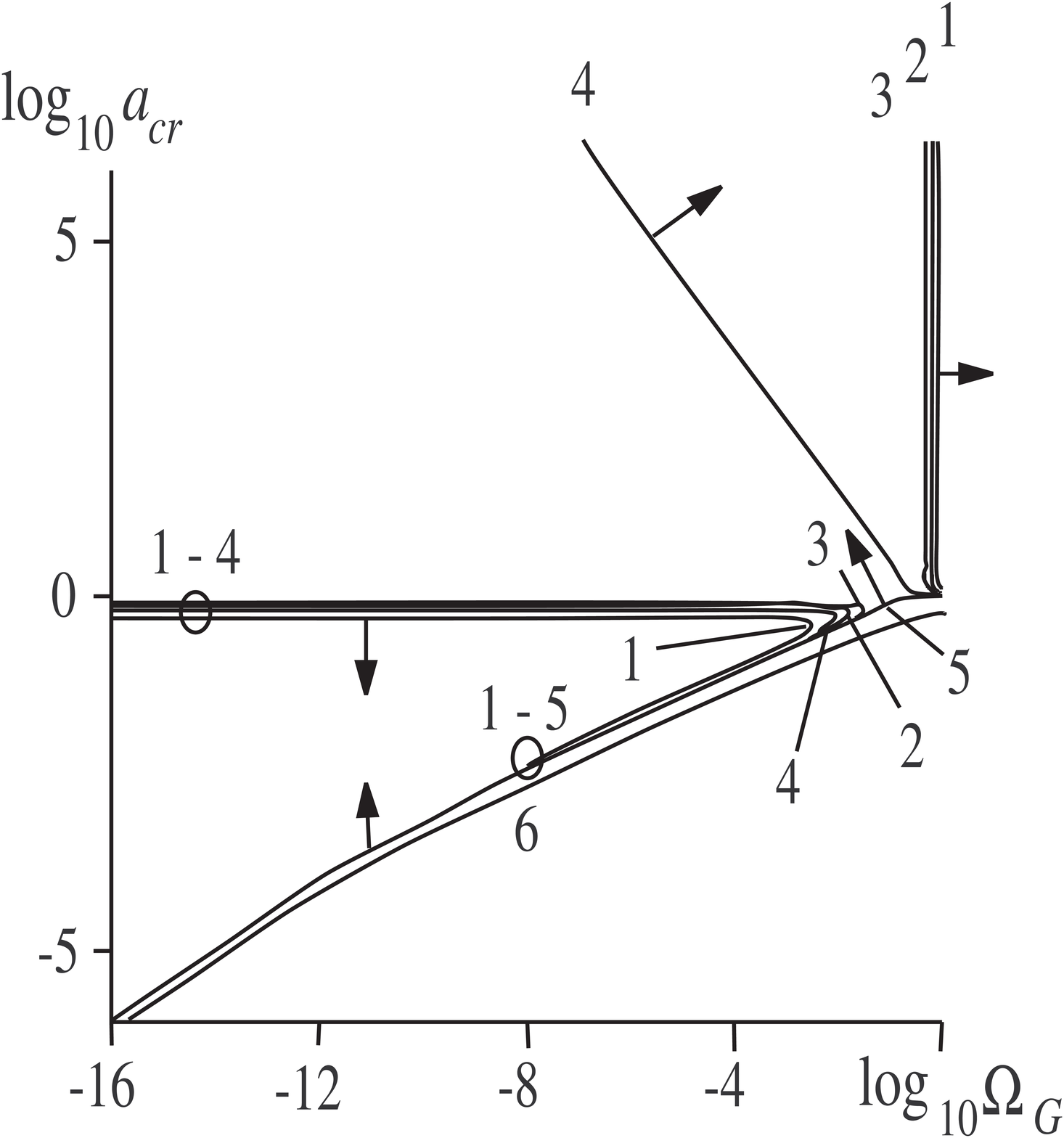,height=7cm} 
   \end{tabular}
   \end{center}
   {Fig. 2. The logarithm of the critical scaling factor in the dependence on the logarithm of the
        graviton's critical density. $\Omega_X$ = 0.8 (1), 0.65 (2, 4, 5), 0.5 (3), $w$ = -1 (1 - 3),
        -2/3 (4), -1/3 (5). Curve 6 shows $a_{min}$, arrows are directed into regions of the
        decelerated expansion} 
   \end{figure} 

It is possible to distinguish following scenarios of the universe's evolution:\\

\indent
$I$. \textit{Deceleration and recollapse}. The initially accelerated expansion changes into
deceleration with subsequent recollapse. Such behavior takes a place for $\Omega_G > \Omega_X$ in
the case of $w$ = -1 (right-hand side group of curves 1 - 3), for $\Omega_G$ lying on the right
of lower curve 4 in the case of $w$ = -2/3 and for all $\Omega_G$ in the case of $w$ = -1/3 (curve 5).\\

\indent
$II$. \textit{Eternal acceleration}. For $w$ = -1 there is the region of the simple accelerated
expansion. This region increases as result of the $\Omega_X$ increase (region between left-hand side
and right-hand side groups of curves 1 - 3).\\

\indent
$III$. \textit{Loitering expansion without recollapse}. This is a complicated scenario existing
for $w$ = -1 and small $\Omega_G$ (left-hand side group of curves 1 - 3). The initially accelerated
expansion changes into deceleration, which changes into an eternal acceleration not long before
the present time. The region of the deceleration increases as result of the $\Omega_X$ decrease
(transition from the left-hand side curve 1 to 3).\\

\indent
$IV$. \textit{Loitering expansion with recollapse}. For $w$ = -2/3 the behavior is the similar to $III$,
but the long acceleration changes into deceleration and recollapse (curves 4). The acceleration can be very
long for small $\Omega_G$.\\

\indent
$V$. \textit{Marginally loitering}. On the right of the region $II$ for $w$ = -1 there is the narrow region
near $\Omega_G = \Omega_X$, where the accelerated expansion changes into decelerated one with the subsequent
transition to eternal acceleration.\\

The examples of the described scenarios are shown in Fig. 3.

 \begin{figure}
   \begin{center}
   \begin{tabular}{c}
   \psfig{figure=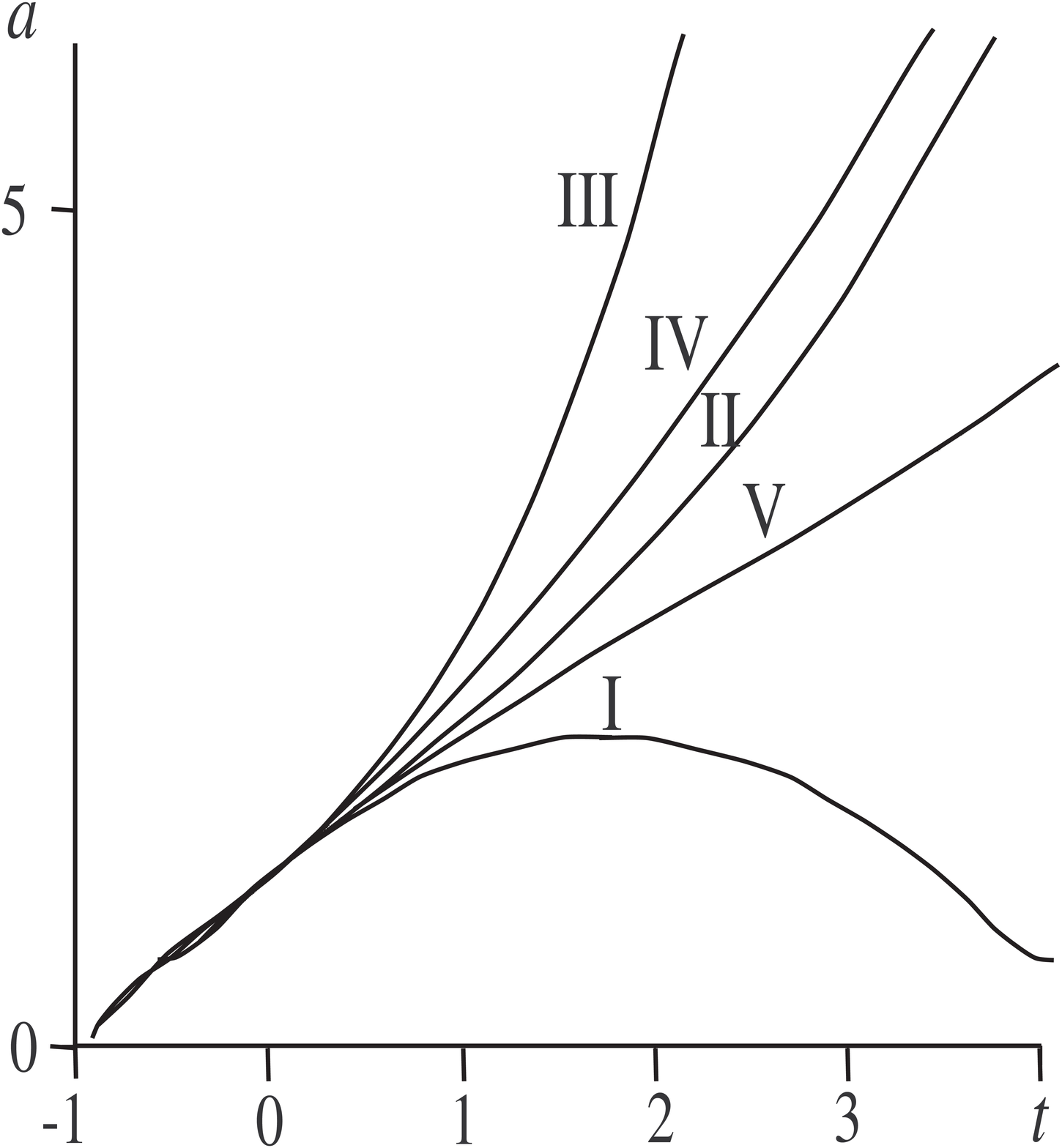,height=7cm} 
   \end{tabular}
   \end{center}
   {Fig. 3. The scenarios I - V. $\Omega_G$ = 0.8 (I), 0.5 (II), $10^{-12}$ (III), $10^{-3}$ (IV), 0.62 (V);
        $w$ = -1 (I - III, V), -2/3 (IV). $\Omega_X$ = 0.65. Recollapse for scenario IV is located outside of Fig.} 
   \end{figure} 

So, in the presence of X-matter there exist the scenarios with acceleration far from $a_{min}$ (II - V). As
the main criterion for the selection of appropriate scenario we consider the demand of the existence of the
radiation dominated epoch. The last restricts the value of $\Omega_G$ so as $a_{min} < 10^{-4} \approx 1/\Omega_R$.
Then from Eq. \ref{9} one can obtain $\Omega_G < 10^{-11.7}$ that gives for the dimensional graviton's
mass $m < 10^{-71} g$ if $H_0 = 68 \textit{km s}^{ - 1} \textit{ Mps}^{ - 1}$. This estimation is essentially
lower than in \cite{4}: $m < 4.5 \ast 10^{-66} g$, because of the last value was obtained on the basis of
criterion $a_{max} > 1$ (i. e. the present epoch corresponds to expansion for $\Omega_X$ = 0). Obviously,
that our criterion is more rigid.

The obtained estimation allows to prefer the scenarios III or IV for the case of the accelerated expansion.
The next step is the calculation of the universe's age, that can be made by the integration of Eq. \ref{6}:

\begin{equation}
\tau _0  = \frac{1}{{H_0 }}\int_{a_{\min } }^1 {\frac{{da}}{{\sqrt A }}} ,
\end{equation}
\noindent
where
\begin{eqnarray*}\label{11}
A = \frac{{1 - \Omega _X  - \Omega _R }}{a} + \frac{{\Omega _X }}{{a^{1 + 3w} }} + \frac{{\Omega _R }}{{a^2 }} -\\
 \Omega _g \left( {a^2  + \frac{1}{{2a^4 }} - \frac{3}{{2a^2 }}} \right).
\end{eqnarray*}

It is convenient to divide the integration on the two part: radiation dominated epoch and matter dominated
epoch. For the first interval one can use the approximation, which produced Eq.\ref{9}. Then the duration
of the radiation dominated epoch is $\sim 10^{5}$ yr with slow dependence on $\Omega_G$ for $a_{min} \ll 1/\Omega_R$.
The further integration for the matter dominated epoch gives the results, which are presented in Table 2
($H_0 = 68 \textit{km s}^{ - 1} \textit{ Mps}^{ - 1}$). The obtained values of the universe's age agree
with the modern observational data (see Table 1) and don't depend on the graviton's mass.

\begin{table}
\caption{Universe's age for scenarios III, IV}
\begin{center}
\begin{tabular}{|c|c|c|}
\hline
 $\Omega_{X}$ & $\tau_{0}(w=-1)$, Gyr & $\tau_{0}(w=-2/3)$, Gyr\\
\hline
			0.5 & 11.6 & 11.2 \\
        0.65 & 12.9 & 12.2 \\
			 0.8 & 15 & 13.8\\        
\hline
\end{tabular}
\end{center}
\end{table}

\section{Conclusion}

The introducing of the $\Lambda$-term in the Lagrangian of RTG demands $\Lambda = -m^2$ that contributes to
the deceleration of the universe's expansion. However, modern observations suggest the accelerated expansion
at the present epoch. We showed that without modification of the field equation \cite{8} 
the insertion of X-matter with
the equation of the state, which is close to the extremal weak energy condition $\rho + p = 0$ extends
the class of the cosmological models in framework of RTG. As result, there exist the
cosmological scenarios, which agree with the observational data. We demonstrated, that in the case of
$m < 10^{-71} g$ and $-1 \leq w \leq -2/3$ the minimal size of the universe, its age and acceleration at
the present epoch don't contradict with the models considered in the framework of GR. But the existing data
don't allow to choose between different scenarios because the similar present behavior results in the
essential difference in the future: eternal expansion for $w = -1$ (or $m = 0$), and recollapse for $w > -1$.
Moreover, the evolution, as rule, behaves the complicated loitering character with alternation of the acceleration
and deceleration. That demands the additional investigations in this direction, especially, the search of RTG
consequences for the universe's structure formation.

This work was carried out by the use of the computer algebra system Maple 6. We suppose, that the algorithmization
of the RTG basics can stimulate the investigations in this direction. The corresponding commented program can be
found and downloaded on \cite{11}.

Author is Lise Meitner fellow at the Technical University of Vienna (project M611).


\small
\begin{thebibliography}{13}

\bibitem{1}
     A.A. Logunov. ``Relativistic theory of gravity'', Nova Sc. Publication,
1998.

\bibitem{2}
		A.A. Logunov, M.A. Mestvirishvili. Relativistic theory of gravitation, Nauka,
Moscow, 1989 (in Russian).

\bibitem{3}
     S.S. Gerstein, A.A. Logunov, M.A. Mestvirishvili, {\it Phys. Atomic Nuclei}, {\bf 61},
1420 (1998).

\bibitem{4}
	S.S. Gerstein, A.A. Logunov, M.A. Mestvirishvili, arXiv: hep-th/9711147 (1997).

\bibitem{5}
	L. M. Krauss, arXiv: astro-ph/0102305 (2001).

\bibitem{6}
	A.G. Riess et al. {\it Astron. Journ.}, {\bf 116}, 1009 (1998).

\bibitem{7}
	S. Permutter et al. {\it Astrophys. Journ.}, {\bf 517}, 565 (1999).

\bibitem{8}
  V.V. Kiselev, arXiv: hep-ph/0106003 (2001).

\bibitem{9}
	S.M. Carroll, {\it Living reviews in relativity}, {\bf 4}(1) (2001).\\
  P. Bin\'etruy, arXiv: hep-ph/0005037 (2000).

\bibitem{10}
	V.L. Kalashnikov, arXiv: gr-qc/0103023 (2001).

\bibitem{11}
	V.L. Kalashnikov, http://www.geocities.com/optomaplev/programs/rtg1.html

\bibitem{12}
	J.D. Cohn, arXiv: astro-ph/9807128 (1998).

\bibitem{13}
	D.H. Lyth, arXiv: astro-ph/9312022 (1993).

\end{thebibliography}
\end{document}